\documentclass[%
preprint,
 amsmath,amssymb,
 aps,
]{revtex4-1}

\usepackage{tikz}
\usepackage{graphicx}
\usepackage{dcolumn}
\usepackage{bm}
\usepackage{hyperref}

\begin{document}


\title{The Travelling Salesman Problem and\\Adiabatic Quantum Computation: An Algorithm }

\author{Tien D. Kieu}
 \email{tien.d.kieu@gmail.com}
\affiliation{%
Centre for Quantum and Optical Science\\
 Swinburne University of Technology, Victoria, Australia
}%




\date{\today}

\begin{abstract}
An explicit algorithm for the travelling salesman problem is constructed in the framework of adiabatic quantum computation, AQC.  The initial Hamiltonian for the AQC process admits canonical coherent states as the ground state, and the target Hamiltonian has the shortest tour as the desirable ground state.  Some estimates/bounds are also given for the computational complexity of the algorithm with particular emphasis on the required energy resources, besides the space and time complexity, for the physical process of (quantum) computation in general.

\end{abstract}

\pacs{Valid PACS appear here}
\maketitle






One of the most well-known problems in combinatorial optimisation and also in computer science is the traveling salesman problem (TSP), whereby a set of cities and distances between each pair of cities are given and whose solution is a path that visits each city exactly once and returns to the starting city, such that the total distance travelled is minimised~\cite{Lawler, Applegate}. The TSP, since its formulation in 1930's, is also important because it is both difficult, $NP$-hard, and representative, $NP$-complete.  $NP$-hard, non-deterministic polynominal hard, problems are computationally intractable in the sense that no known solutions in polynomial time, as functions of the sizes of the inputs, have been found -- as distinct from the class of $P$ problems which can be solved in exactly polynominal time.  As a subset of $NP$ problems, the $NP$-complete problems are those whose solutions are sufficient to deal with any other $NP$ problems in polynomial time.  Thus, if efficient solutions could be found for any $NP$-complete problem, including the TSP, then we would be able to resolve positively the unanswered question whether $P \stackrel{?}{=} NP$.

On the other hand, quantum computation has been demonstrated to be superior to classical Turing computation for a handful of specific problems~\cite{Nielsen, 1611.04471}.  
For the TSP, several quantum algorithms have been attempted~\cite{Moser, Talbi, Martonak, Moylett, Heim, 0411013}.  
We propose in the present paper another explicit quantum algorithm in the framework of Adiabatic Quantum Computation (AQC).  Some initial seeds for the results herein have been mentioned and speculated in an earlier paper~\cite{Kieu}.  AQC is based on the obtainability of some quantum ground states but in general it is not straight forward to rephrase an optimisation problem in terms of ground states of some suitably constructed quantum Hamiltonians.  Of the few other AQC algorithms available thus far is, for example, one for the problem of factorising an integer into its prime constituents~\cite{kieu_factorisation}.  It would be interesting if AQC algorithms could be found for more $NP$-hard problems, for instance the class discussed in~\cite{parameterized_algos}.

We briefly summarise the principles of AQC computation in the next section.  Followed that is the explicit construction of our algorithm.  We then discuss next the computational complexity of the proposal with strong emphasis on the energy, as well as the time and space, required for any physical computation.  Our complexity arguments and speculations are illustrated further and supported by some well-known AQC algorithms in Appendix B.  These considerations require some recent results, which are summarised in Appendix A, on a new class of uncertainly relations of time and energy for general time-dependent quantum Hamiltonians~\cite{Kieu2}.  Our time-energy consideration points to a dedicate balance between the lower bound of the time limit and the required energy resources to carry out the computation.  It is perhaps not that surprising that we may be able to reduce the computation time with more energy resource, as the TSP is well parallelisable.  For example, in a parallel computation of $(N-1)!$ processing units, one for each of the permutations of $(2, \ldots, N)$ cities, we could solve the problem in efficient time but at the cost of a huge number of computer processors.  One of the advantages of our proposed AQC algorithm is the existence of arbitrary $c$-number parameters, of which suitable choices would ensure some optimal balance between the lower bound on the time and the energy required for computation.

\section{\label{sec:level2}Adiabatic Quantum Computation (AQC)}
AQC starts with the readily constructible ground state $|g_I\rangle$
of an initial Hamiltonian $H_I$ which is then {\em adiabatically}
extrapolated to the final Hamiltonian $H_P$ whose ground state
$|g_P\rangle$ encodes the desirable
solution of the problem and could be then obtained with
reasonably high probability.	The interpolation between $H_I$ and
$H_P$ is facilitated by a time-dependent Hamiltonian in the time
interval $0\le t \le T$,
\begin{eqnarray}
{\cal H}(t) &=& f(t/T)H_I + g(t/T)H_P,
\label{extrapolation}
\end{eqnarray}
either in a temporally linear manner (that is, $f(t/T) =
\left(1-{t}/{T}\right)$ and $g(t/T)={t}/{T} $); or otherwise with $f+g=1$, and $f(0) = 1 = g(1)$ and $f(1) = 0 = g(0)$.  We also assume that both $f$ and $g$ are continuous, and $g$ is semi-positive for all $t\in[0,T]$.	Such a time
evolution is captured by the Schr\"odinger equation:
\begin{eqnarray}
i\partial_t |\psi(t)\rangle &=& {\cal H}(t)\;|\psi(t)\rangle,
\label{AQC}\\
|\psi(0)\rangle &=& |g_I\rangle. \nonumber \label{A5}
\end{eqnarray}

\section{\label{sec:level3}The initial Hamiltonian and its Ground State for the TSP}
We will require three sets of creation-annihilation operators, one set for the links between each pair of cities, one set of `hookers' for each city, and one set of `markers' for each city.  Associated with each set is the space of occupation-number states, and the full state space is the direct product of these separate spaces.

We introduce the link operators $l^\dagger_{ij}$ and $l_{ij}$, which respectively creates and annihilates a link emanating from the $j$-th and ending at the $i$-th cities:
\begin{eqnarray}
\left[l_{ij}, l^\dagger_{i'j'}\right] &=& \delta_{ii'}\delta_{jj'},
\end{eqnarray}
where all the indices range from 1 to $N$.  The link occupation states $|n\rangle_{l_{ij}}$ ($n=0,1,2,\ldots$), and the number operators $\hat{n}_{ij}$ are constructed in the usual way:
\begin{eqnarray}
l^\dagger_{ij} |n\rangle_{l_{ij}} &=& \sqrt{n+1}|n+1\rangle_{l_{ij}}, \nonumber\\
l_{ij} |n\rangle_{l_{ij}} &=& \sqrt{n}|n-1\rangle_{l_{ij}}, (n\ge 1),\nonumber\\
l_{ij} |0\rangle_{l_{ij}} &=& 0,\nonumber\\
\hat{n}_{ij} &=& l^\dagger_{ij}l_{ij}.
\end{eqnarray}

We next introduce the ``hooker" operators $h^\dagger_i$ (creation) and $h_j$ (annihilation), with $i$, $j = 1\ldots N$, together with the hooker occupation state $|n\rangle_{h_i}$:
\begin{eqnarray}
\left[h_i, h^\dagger_j\right] &=& \delta_{ij}, \nonumber\\
h_i|0\rangle_{h_i} &=& 0.
\end{eqnarray}

We also require the ``marker" operators $m^\dagger_i$ (creation) and $m_j$ (annihilation), with $i$, $j = 1\ldots N$, together with the marker occupation state $|n\rangle_{m_i}$:
\begin{eqnarray}
\left[m_i, m^\dagger_j\right] &=& \delta_{ij}, \nonumber\\
m_i|0\rangle_{m_i} &=& 0.
\end{eqnarray}

Otherwise, all the operators in the sets $\{l\}$, $\{h\}$ and $\{m\}$ are pairwise commuting across the sets.

We could now start the AQC with the following initial Hamiltonian $H_I$:
\begin{eqnarray}
 H_I &=& \sum_{ij} (l^\dagger_{ij} - \theta_{ij}^*)(l_{ij} - \theta_{ij})
+ \sum_{i}h^\dagger_i h_i + \sum_{i}m^\dagger_i m_i, \label{H_I}
\end{eqnarray}
for some c-numbers $\theta_{ij}$.  This Hamiltonian admits the readily constructible ground state $|g_I\rangle$:
\begin{eqnarray}
|g_I\rangle &=& |\{\theta\}\rangle_{\{l\}} \otimes |0\rangle_{\{h\}} \otimes |0\rangle_{\{m\}}, \label{7}
\end{eqnarray}
with $|0\rangle_{\{h\}}$ and $|0\rangle_{\{m\}}$ are correspondingly the cross products of the zero-number states for the operators $h$ and $m$:
\begin{eqnarray}
|0\rangle_{\{h\}} &=& \bigotimes_{i\not=1}^N |0\rangle_{h_i},\\
|0\rangle_{\{m\}} &=& \bigotimes_{i\not=1}^N |0\rangle_{m_i},
\end{eqnarray}
and $|\{\theta\}\rangle_{\{l\}} $ is the cross product of the canonical quantum coherent states $|\theta_{ij}\rangle$ for the operators $l_{ij}$: 
\begin{eqnarray}
|\{\theta\}\rangle_{\{l\}}  &=& \bigotimes_{ij}|\theta_{ij}\rangle, \\
|\theta_{ij}\rangle &=& {\rm e}^{-{\frac{|\theta_{ij}|^2}{2}}}\sum_{n=0}^\infty \frac{(\theta_{ij})^{n}}{\sqrt[]{n!}}|n\rangle_{l_{ij}}, \nonumber
\end{eqnarray}
where:
\begin{eqnarray} 
l_{ij} \; |\theta_{ij}\rangle &=& \theta_{ij}\;|\theta_{ij}\rangle. \nonumber
\end{eqnarray}
The ground state $|g_I\rangle$ of~(\ref{7}) thus contains the products of all the possible pairwise links between the cities, corresponding to traversed links (when $n_{ij}\not=0$) that could have more than one traversal ($n_{ij}>1$).

\section{Filtering Operators for Connected and Complete Tours}
All the paths are encoded in the number states of the link operators $l$,
\begin{eqnarray}
|{\rm paths}\rangle &=& |\{n\}\rangle_{\{l\}}.
\end{eqnarray}
The paths could be segmented, or form tours (closed loops) or subtours.  For example, with 4 cities, the state $|1_{23}2_{14}\{0_{ij}\}\rangle$, where $(i,j) \not= (2,3)$ and $(i,j) \not= (1,4)$, encodes a path with two segments, one segment connects cities 2 and 3, and the other connects cities 1 and 4 but with two traversals.  And the state $|1_{12}1_{23}1_{34}1_{41}\{0 _{ij}\}\rangle$, with $(i,j)\not\in\{(1,2),(2,3),(3,4),(4,1)\}$, encodes a tour, which is the closed path 1-2-3-4-1, with each link traversed only once.

Note that the number of traversals on a link could be zero, one, or more than one.  The number operator $\hat{n}_{ij}$ counts the number of traversals on the link between the $i$-th and $j$-th cities.  We define a {\em connected} tour as the case when there is no disjoint subtours, and {\em complete} tour as when all the cities are included.

For a link starting from $i$ and ending with $j$ (that is, $n_{ji}\not =0$) we employ the hooker operator $h_j^\dagger$ to create a ``hook" at $j$.  We now construct the first layer operator $F$ for $(N-1)$ links emanating from the starting city $i=1$ to all other cities in such a way that a hook is created at each end point of each link:
\begin{eqnarray}
F &=& \sum^N_{j} m^\dagger_j h_j^\dagger \hat{n}_{j1},
\end{eqnarray}
the requirement for $m^\dagger_j$ to create a ``marker" and the end point of each link will soon be evident in the below.

\begin{figure}[h!]
  \begin{center}
    \begin{tikzpicture}
      \draw [red,fill] (-2,-2) circle [radius=0.2] node [black,below=4] {$i$};
      
      \draw [thick, ->] (-1.7, -2) 
      to [out=10,in=180] (1,0.5);
      \draw [thick, ->] (1,-2) 
      to [out=-170,in=-90] (-2,-2.6);
      
      
      

  \label{f1}
    \end{tikzpicture}
    \caption{A link emanating from $i$ carries a hooker annihilation $h_i$ cancelling the effect of the hooker creation $h^\dagger_i$ from another arriving link at $i$, provided the arriving link can be traced back to the starting city labelled {\bf 1}, in which case this arriving link also creates a marker  at $i$ by $m^\dagger_i$.}    
  \label{fig:f1}
  \end{center}
\end{figure}
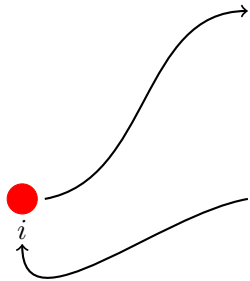

For the next layers of $N$ links in a tour, we construct an operator $L$ such that a hook is destroyed at the emanating end of each link and another hook created at the arriving end, see FIG.~\ref{fig:f1}:
\begin{eqnarray}
L &=& \sum_{ ij} m^\dagger_j h^\dagger_j \hat{n}_{ji} h_i.
\end{eqnarray}
To close up the tours we finally introduce the ending layer operator $E$ for $(N-1)$ links arriving at the city $i=1$:
\begin{eqnarray}
E = \sum^N_{j} \hat{n}_{1j} h_j,
\end{eqnarray}
where we still require the hook annihilation operators $h_j$ at the emanating end of each link.

The combination operator $Q$,
\begin{eqnarray}
Q &=& \left(\prod^N_{i\not=1} m_i\right)E\left(L\right)^{(N-2)}F,
\end{eqnarray}
with $(N-2)$ factors of $L$, admits the following states as eigenvectors, with integers $n$, $p$ and $q$,
\begin{eqnarray}
|\{n\}\rangle_{\{l\}}\otimes |\{p\}\rangle_{\{h\}}\otimes |\{q\}\rangle_{\{m\}}. \label{16}
\end{eqnarray}
This is because $Q$ only contains the number operators $\{\hat{n}_{ij}\}$ and balanced numbers of creation and annihilation operators in $\{h^\dagger_i,h_i\}$ and also separately in $\{m^\dagger_i,m_i\}$.

Since the states (\ref{16}) having non-zero occupation numbers in $h$ and $m$, that is $|n_i\not=0\rangle_{\{h\}}\otimes|n_j\not=0\rangle_{\{m\}}$ for some $i$ and $j$, they may only contribute to states of higher eigenvalues of the target Hamiltonian of the AQC to be introduced later on in (\ref{H_P}).  From now on it suffices, for the purpose of our AQC, to consider the following paths only:
\begin{eqnarray}
|\{n\}\rangle_{\{l\}}\otimes |0\rangle_{\{h\}}\otimes |0\rangle_{\{m\}}.
\end{eqnarray}

The role of the hookers $\{h^\dagger_i,h_i\}$ in $Q$ is to ensure that $Q$ will preserve only the states having only one arriving and one departing link from each intermediate city that can be traced back to the starting city 1.  The hookers will ``hook" those links into closed tours, includding subtours, going through the city $i=1$ and having exactly $N$ traversed links.

The other factor $\left(\prod^N_{i\not=1} m_i\right)$ in $Q$ is there to enforce that these connected tours are indeed {\em complete} in the sense that they visit {\em all} the $N$ cities.  This is because if an intermediate city $k$ is not visited, i.e. $\hat{n}_{kj}=0$, for all $j$, then there will be no action of the creator $m^\dagger_k$ on $|0\rangle_{m_k}$, and thus  $|0\rangle_{m_k}$ will be annihilated by the action of the annihilator $m_k$ in the factor $\left(\prod^N_{i\not=1} m_i\right)$.  On the other hand, the state $|p\not =0\rangle_{m_k}$ is not an eigenstate of $m^\dagger_k$.

{\em All} other tour configurations are either annihilated or not admitted as eigenstates of the operator $Q$.

Some configurations $|{\rm paths}\rangle|0\rangle_{h}|0\rangle_{m}$ that are not survived by the action of $Q$ are illustrated in the figures below for the case of four cities, $N=4$.  

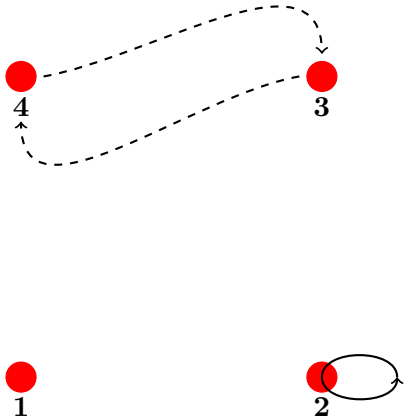
\begin{figure}[h!]
  \begin{center}
    \begin{tikzpicture}
      \draw [red,fill] (-2,-2) circle [radius=0.2] node [black,below=4] {\bf 1};
      \draw [red,fill] (2,-2) circle [radius=0.2] node [black,below=4] {\bf 2}; 
      \draw [red,fill] (2,2) circle [radius=0.2] node [black,below=4] {\bf 3};
      \draw [red,fill] (-2,2) circle [radius=0.2] node [black,below=4] {\bf 4};
      
      
      \draw [thick, dashed, ->] (-1.7,2) 
      to [out=10,in=90] (2,2.3);      
      \draw [thick, dashed, ->] (1.7,2) 
      to [out=-170,in=-90] (-2,1.4);
      
      
      \draw [thick, ->] (2,-2) 
      to [out=-90,in=-90] (3,-2);
      \draw [thick] (3,-2) 
      to [out=90,in=90] (2,-2);

    \end{tikzpicture}
    \caption{No tour starts from {\bf 1}.  (The continous link indicates that the link occupation number is not zero and there is a corresponding link operator from $Q$ acting on this particular link.  The dotted links also indicate that the link occupation numbers are non-zero, but there is no corresponding link operator from $Q$.)}   
  \label{fig:f2}
  \end{center}
\end{figure}

The configuration state in FIG.~\ref{fig:f2} does not contribute because there is no link starting from the city 1 and hence is eliminated by the factor $F$ in $Q$.

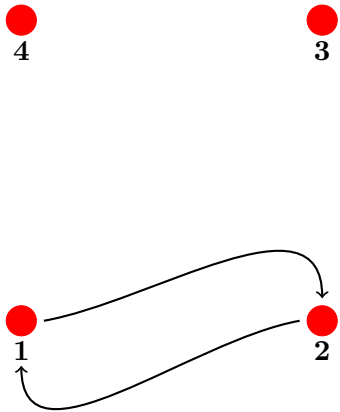
\begin{figure}[h!]
  \begin{center}
    \begin{tikzpicture}
      \draw [red,fill] (-2,-2) circle [radius=0.2] node [black,below=4] {\bf 1};
      \draw [red,fill] (2,-2) circle [radius=0.2] node [black,below=4] {\bf 2}; 
      \draw [red,fill] (2,2) circle [radius=0.2] node [black,below=4] {\bf 3};
      \draw [red,fill] (-2,2) circle [radius=0.2] node [black,below=4] {\bf 4};
      
      \draw [thick, ->] (-1.7, -2) 
      to [out=10,in=90] (2,-1.7);
      \draw [thick, ->] (1.7,-2) 
      to [out=-170,in=-90] (-2,-2.6);
      
      
      

    \end{tikzpicture}
    \caption{Tour starts from {\bf 1} with less than $N$ traversed links.}   
  \label{fig:f3}
  \end{center}
\end{figure}

The configuration state in FIG.~\ref{fig:f3} is eliminated because the tour, even though starts from city 1, has less than 4 traversed links that can be traced back to 1, and hence does not match the number of link factors in $Q$.

\begin{figure}[h!]
  \begin{center}
    \begin{tikzpicture}
      \draw [red,fill] (-2,-2) circle [radius=0.2] node [black,below=4] {\bf 1};
      \draw [red,fill] (2,-2) circle [radius=0.2] node [black,below=4] {\bf 2}; 
      \draw [red,fill] (2,2) circle [radius=0.2] node [black,below=4] {\bf 3};
      \draw [red,fill] (-2,2) circle [radius=0.2] node [black,below=4] {\bf 4};
      
      \draw [thick, ->] (-1.7, -2) 
      to [out=10,in=90] (2,-1.7);
      \draw [thick, ->] (1.7,-2) 
      to [out=-170,in=-90] (-2,-2.6);
      
      
      \draw [thick, ->] (-2,-2) 
      to [out=-90,in=-90] (-3,-2);
      \draw [thick] (-3,-2) 
      to [out=90,in=90] (-2,-2);
      
      \draw [thick, ->] (2,-2) 
      to [out=-90,in=-90] (3,-2);
      \draw [thick] (3,-2) 
      to [out=90,in=90] (2,-2);

    \end{tikzpicture}
    \caption{Incomplete tour with cities {\bf 3} and {\bf 4} not visited.}   
  \label{fig:f4}
  \end{center}
\end{figure}
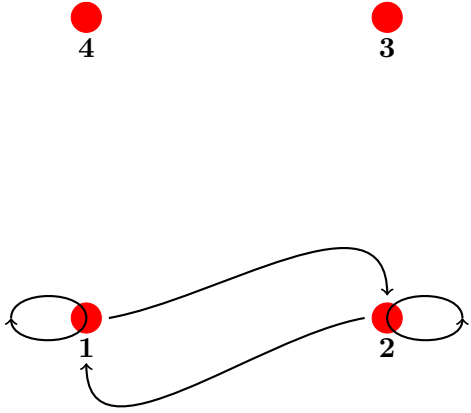

The configuration state in FIG.~\ref{fig:f4} has unvisited cities and is thus eliminated by the marker annihilators $m_i$ (where $i$ is the unvisited city) when $Q$ acting on the state $|0\rangle_{m_i}$.

\begin{figure}[h!]
  \begin{center}
    \begin{tikzpicture}
      \draw [red,fill] (-2,-2) circle [radius=0.2] node [black,below=4] {\bf 1};
      \draw [red,fill] (2,-2) circle [radius=0.2] node [black,below=4] {\bf 2}; 
      \draw [red,fill] (2,2) circle [radius=0.2] node [black,below=4] {\bf 3};
      \draw [red,fill] (-2,2) circle [radius=0.2] node [black,below=4] {\bf 4};
      
      \draw [thick, ->] (-1.7, -2) 
      to [out=10,in=90] (2,-1.7);
      \draw [thick, ->] (1.7,-2) 
      to [out=-170,in=-90] (-2,-2.6);
      
      \draw [thick, ->] (2.3, -1.7) 
      to [out=80,in=0] (2.3, 2);
      \draw [thick, ->] (1.7,1.7) 
      to [out=-100,in=-180] (1.6,-1.9);
      
      
      

    \end{tikzpicture}
    \caption{Incomplete tour with city {\bf 4} not visited.}   
  \label{fig:f4a}
  \end{center}
\end{figure}
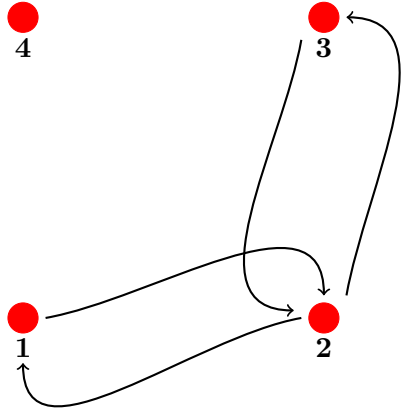

Same situation for FIG.~\ref{fig:f4a}.  In general, when a site is not visited (like site {4} here) or visited more than once (like site {2} here) there will be an imbalance between $m$ and $m^\dagger$ in $Q$, eliminating that configuration state out of consideration.

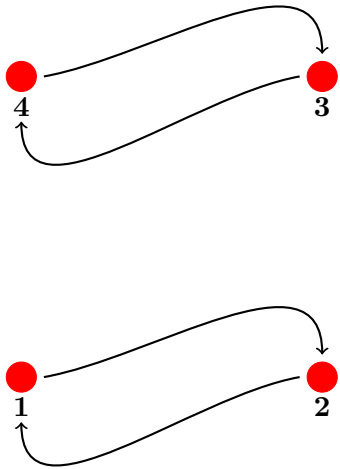
\begin{figure}[h!]
  \begin{center}
    \begin{tikzpicture}
      \draw [red,fill] (-2,-2) circle [radius=0.2] node [black,below=4] {\bf 1};
      \draw [red,fill] (2,-2) circle [radius=0.2] node [black,below=4] {\bf 2}; 
      \draw [red,fill] (2,2) circle [radius=0.2] node [black,below=4] {\bf 3};
      \draw [red,fill] (-2,2) circle [radius=0.2] node [black,below=4] {\bf 4};
      
      \draw [thick, ->] (-1.7, -2) 
      to [out=10,in=90] (2,-1.7);
      \draw [thick, ->] (1.7,-2) 
      to [out=-170,in=-90] (-2,-2.6);
      
      \draw [thick, ->] (-1.7,2) 
      to [out=10,in=90] (2,2.3);      
      \draw [thick, ->] (1.7,2) 
      to [out=-170,in=-90] (-2,1.4);
      
      

    \end{tikzpicture}
    \caption{Complete but having disjoint subtours.}   
  \label{fig:f5}
  \end{center}
\end{figure}

The configuration state in FIG.~\ref{fig:f5}, even though having all the cities visited but with {\em disjoint} subtours as there are cities (3 and 4) not connected to city 1, cannot contribute because it is eliminated by the hooker annihilators $h_i$ in $L$ acting on $|0\rangle_{h_i}$, with $i=3$ or $4$.  It would survive only if the $h$-occupation number state at site 3 or 4 is not zero, namely  $|p\not=0\rangle_{h_3}$ or $|p\not=0\rangle_{h_4}$.  However, we will construct the final AQC Hamiltonian as in~(\ref{H_P}) below in such a way that these non-zero occupation states will not be the final ground state.

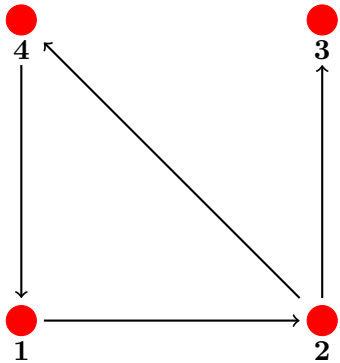
\begin{figure}[h!]
  \begin{center}
    \begin{tikzpicture}
      \draw [red,fill] (-2,-2) circle [radius=0.2] node [black,below=4] {\bf 1};
      \draw [red,fill] (2,-2) circle [radius=0.2] node [black,below=4] {\bf 2}; 
      \draw [red,fill] (2,2) circle [radius=0.2] node [black,below=4] {\bf 3};
      \draw [red,fill] (-2,2) circle [radius=0.2] node [black,below=4] {\bf 4};
      
      \draw [thick, ->] (-1.7,-2) 
      to [] (1.7,-2);
      
      \draw [thick, ->] (2,-1.7) 
      to [] (2,1.4);      
      \draw [thick, ->] (1.7,-1.7) 
      to [] (-1.7,1.7);
      
      \draw [thick, ->] (-2,1.4) 
      to [] (-2,-1.7);

    \end{tikzpicture}
    \caption{A broken tour which has only one hooker creation $h^\dagger_2$ from the {\bf 1-2} link arriving at {\bf 2}, not balancing with two hooker annihilators $h_2$ from the {\bf 2-3} and {\bf 2-4} links emanating from {\bf 2}.}   
  \label{fig:f6}
  \end{center}
\end{figure}

FIG.~\ref{fig:f6} depicts a broken tour that does not survive the action of $Q$ because the number of hooker creators from the arriving link at the city 2 does not match the number of hooker annihilators from the departing links at the same city.  (Remember that the number states $|p\not = 0\rangle_{h_i}$ are not eigenstates of $Q$ when there is an imbalance in the operators $h_i$ and $h^\dagger_i$.)

      
      
      


The operator $Q$ ends up preserving configuration states that contain complete tours with or without subtours, as in FIGs.~\ref{fig:f7} and~\ref{fig:f8}.  Note that for the survived configurations, beside those links making up the desired tours having $n_{ij}=1$, we also have $n_{ij}>1$ but these will not contribute to the lowest eigenvalue 1 of $Q$.  

In fact, 
{\texttt{
\begin{eqnarray}
&&Q\; |{\rm connected\; and\; complete\; tour}; \{n\}_{\rm others}\rangle|0\rangle_{\{h\}} |0\rangle_{\{m\}} = \nonumber\\
&&\left(\prod_{\stackrel{n_{ij}\in {\rm tour}}{n_{ij}\not=0}}n_{ij}\right)|{\rm connected\; and\; complete\; tour}; \{n\}_{\rm others}\rangle|0\rangle_{\{h\}} |0\rangle_{\{m\}}. \label{others}
\end{eqnarray}}}
Thus the states with any $n_{ij}>1$ (with $n_{ij}\in$ tour) will result in higher eigenvalues for $Q$.  
Note that in general the link number state $|\{n\}_{ij}\rangle$, which has  $N^2$ values of $n$ (for $i,j=1,\ldots,N$), there are also links that are not acted upon by components of $Q$, because each term of $Q$ is a product of only $N$ link operators.  We have thus denoted $\{n\}_{\rm others}$ as the set of those links that are {\rm not} acted upon by the link operators in components of $Q$.  These other links are depicted as dotted lines the figures, and they contribute only to the higher eigenstates and not the ground state of the final Hamiltonian~(\ref{H_P}).
\begin{figure}[h!]
  \begin{center}
    \begin{tikzpicture}
      \draw [red,fill] (-2,-2) circle [radius=0.2] node [black,below=4] {\bf 1};
      \draw [red,fill] (2,-2) circle [radius=0.2] node [black,below=4] {\bf 2}; 
      \draw [red,fill] (2,2) circle [radius=0.2] node [black,below=4] {\bf 3};
      \draw [red,fill] (-2,2) circle [radius=0.2] node [black,below=4] {\bf 4};
      
      \draw [thick, ->] (-1.7,-2) 
      to [] (1.7,-2);
      
      \draw [thick, ->] (1.7,-1.7) 
      to [] (-1.7,1.7);      
      \draw [thick, ->] (1.7,1.7) 
      to [] (-1.7,-1.7);
      
      \draw [thick, ->] (-1.7,2) 
      to [] (1.7,2);
      
      \draw [thick, dashed] (2,-1.7) 
      to [] (2,1.3);
      
       \draw [thick, dashed] (-2,1.3) 
      to [] (-2,-1.7);

    \end{tikzpicture}
    \caption{A connected and complete tour.}   
  \label{fig:f7}
  \end{center}
  \begin{center}
    \begin{tikzpicture}
      \draw [red,fill] (-2,-2) circle [radius=0.2] node [black,below=4] {\bf 1};
      \draw [red,fill] (2,-2) circle [radius=0.2] node [black,below=4] {\bf 2}; 
      \draw [red,fill] (2,2) circle [radius=0.2] node [black,below=4] {\bf 3};
      \draw [red,fill] (-2,2) circle [radius=0.2] node [black,below=4] {\bf 4};
      
      \draw [thick, ->] (-1.7,-2) 
      to [] (1.7,-2);
      
      \draw [thick, ->] (2,-1.7) 
      to [] (2,1.4);      
      \draw [thick, ->] (1.7,2) 
      to [] (-1.7,2);
      
      \draw [thick, ->] (-2,1.4) 
      to [] (-2,-1.7);

    \end{tikzpicture}
    \caption{Shortest tour when all traversed links have $n_{ij} = 1$.}   
  \label{fig:f8}
  \end{center}
\end{figure}

It then follows that the hermitean operator $\left(Q^\dagger-1\right)\left(Q-1\right)$
 {\em annihilates} all the complete tours (with or without subtours)
where each traversed link $ij$, that corresponds to $\hat{n}_{ij}$ in a component of $Q$, has exactly ${n}_{ij}=1$.  Besides these acted upon links, the subset of $\{n_{\rm others}\}$ can have any occupation numbers, but those with non-zero values will not be the ground state of~(\ref{H_P}) below.  The same effect could also be achieved by the operator $\left(Q-1\right)^2$.

\section{\label{sec:level2}The Target Hamiltonian and its Ground State for the TSP}
For the $N$ cities in the TSP, we denote $d_{ij}$ as the given distances between the $i$-th and $j$-th cities, with $d_{ii}=0$ for any $i$.  We can now explicitly construct the target hermitean Hamiltonian $H_P$ for our AQC algorithm:
\begin{eqnarray}
H_P &=& s\left(\sum_i h^\dagger_i h_i + \sum_i m^\dagger_i m_i\right) \nonumber\\
&&+s\left(Q^\dagger-1\right)\left(Q-1\right) \nonumber\\
&& + \sum_{ij} d_{ij} \hat{n}_{ij}, \label{H_P}
\end{eqnarray}
where the factor $s$ is chosen to be greater than the length of the desirable solution tour,
\begin{eqnarray}
s &=& \frac{1}{2}\sum_{ij} d_{ij}. \label{1}
\end{eqnarray}
With that choice, the first two terms on the rhs are to ensure that the ground state of $H_P$ has the factor $|0\rangle_{\{h\}} |0\rangle_{\{m\}}$.  
The next term is to penalise all the paths that are neither connected nor complete.  They also penalise connected and complete tours which have some links in the tours that have $n_{ij}>1$.  The last remaining term measures the lengths of the tours.

This last term in~(\ref{H_P}) is also crucial to ensure that the target ground state has zero occupation number in these other links, $\{n\}_{\rm others} = 0$, in~(\ref{others}).  Otherwise, if there is any non-zero occupation value in any of these other links, the last term of~(\ref{H_P}) would give the state $|{\rm connected\; and\; complete\; tour}; \{n\not = 0\}_{\rm others}\rangle|0\rangle_{\{h\}} |0\rangle_{\{m\}}$ a higher eigenvalue, thus making it an excited state.

The role of the operators $h$ and $h^\dagger$ is to favour closed-loop tours; while that of $m$ and $m^\dagger$ is to enforce that such tours visit each and every site once and only once.  These tours may have disjoint subtours (as in Fig.~\ref{fig:f5}) but they must have some non-zero $h$-occupation numbers.  Such configuration states, however, cannot be the ground state of~(\ref{H_P}) because of the first term on the rhs of~(\ref{H_P}). 


From the above properties the Hamiltonian $H_P$ clearly has as its ground state $|g_P\rangle$ the desirable state for our AQC, as in FIG.~\ref{fig:f8},
\begin{eqnarray}
|g_P\rangle &=& |{\rm the\; shortest\; connected\; and\; complete\; tour}; \{n\}_{\rm others} = 0\rangle|0\rangle_{\{h\}} |0\rangle_{\{m\}}.
\end{eqnarray}

This ground state is at least doubly degenerate because of the two opposite traversing directions.  We could infer from this ground state the desirable tour and its length for our TSP.  

We could also replace in $H_P$ in expression~(\ref{H_P})
\begin{eqnarray}
s\left(Q^\dagger-1\right)\left(Q-1\right) &\longrightarrow& s\left(Q-1\right)^2 + {\rm h.c.},
\end{eqnarray}
and still have the same desirable ground state, but perhaps with some different time complexity, see below.

Note that we may reduce the number of operators in the AQC somewhat by demanding that $l_{ij}=l_{ji}$ and by eliminating the self-linking operators $l_{ii}=0$.  By suitably rearranging $L$ and $H_P$ we could also remove the need for the hooker and marker, $h_1$ and $m_1$, for the starting city 1.

\section{Computational Complexity}
The computational complexity of an algorithm tells us the computational resources required as a function of the input size of the problem to be solved.  Our AQC algorithm for the TSP here requires, first of all, the resources of O($N^2$) link operators, O($N$) hooker operators and O($N$) marker operators.  They may be available in the framework of quantum optics or otherwise.

The computational time complexity is the required time $T$ as a function of the number of the cities $N$ so that the target ground state $|g_P\rangle$ can be obtained with some desirable probability.  Such time, for a given probability, is determined according to the quantum adiabatic theorem~\cite{Messiah} by the inverse of the size of the smallest energy gap that separates the ground state from the first excited state in the spectral flow of ${\cal H}(t)$ for $t \in [0,T]$.  A separation that does not shrink exponentially in $N$ would provide a quantum  algorithm that could solve the $NP$-hard TSP in polynomial time.  
The time complexity remains an open question, to be determined rigorously from the quantum adiabatic theorem, as an open question.  Resolving such a question is a nontrivial task, requiring some estimate of the minimum size of the energy gap in the spectral flow between the ground state and the first excited state --the smaller the gap the longer the adiabatic evolution required for a reasonable probability to obtain the final solution ground state.  However, we will now present some order-of-magnitude estimate for a lower bound on this time.

We have presented elsewhere~\cite{Kieu2}, and see Appendix A, a necessary condition for the time required in a general AQC for an initial state to  evolve into an orthogonal state under the dynamics of ${\cal H}$.  Namely, it is necessary that the evolution time cannot be less than $T_\perp$,
\begin{eqnarray}
T_\perp &\sim& {\cal O} \left(\frac{1}{\Delta_I E_P\int_0^1 g(\tau) d\tau}\right), \label{time1}
\end{eqnarray}
where $\Delta_I E_P$ is the energy spread of the initial state $|g_I\rangle$ in terms of the target Hamiltonian $H_P$,
\begin{eqnarray}
\Delta_I E_P &\equiv& \sqrt{\langle g_I| H_P^2 |g_I\rangle - \langle g_I| H_P |g_I\rangle^2}.
\end{eqnarray}
It is important to note that only the initial eigenstate $|g_I\rangle$, and neither  the instantaneous eigenstates nor the full time-dependent wave function at any other times, is required for the time condition~(\ref{time1}).  This hallmark of our results in~\cite{Kieu2} enables their wider applicability and usefulness.

Here we crucially observe and will exploit the fact that $T_\perp$ is a function of the parameters $s$ and particularly of $|\theta|$,
\begin{eqnarray}
T_\perp &=& T_\perp(s, |\theta|),
\end{eqnarray}
in order to {\rm reduce} the lower time limit $T_\perp$ for the computation as in the below.


From the properties of the coherent state $|\theta\rangle$ 
 we could estimate for large $N$ and small $|\theta|\ll 1$
\begin{eqnarray}
\Delta_I E_P &\sim&  {\cal O}\left(s\sqrt{(N-1)!}|\theta|^N\right), \label{23}
\end{eqnarray}
here we have set $\theta_{ij} = \theta$, for all $i$, $j$.  The factor $(N-1)!$ comes from the enumeration of all the number of complete tours starting from the city 1 and visiting each other city only once.  The exponent $N$ comes from the number of traversed links in a connected and complete tour.  

On the other hand, setting up the initial state $|g_I\rangle \sim \otimes |\theta\rangle$ would require an energy,  for large $N$ and small $|\theta|\ll 1$,
\begin{eqnarray}
\max_{0\le t \le T} \langle g_I|{\cal H}(t)|g_I\rangle &\sim& \left(\int_0^1 g(\tau) d\tau\right){\cal O}\left(s(N-1)!|\theta|^{2N}\right). \label{24}
\end{eqnarray}

As the lower time limit $T_\perp$ is inversely proportional to $\Delta_I E_P$ according to~(\ref{time1}), this lower time limit $T_\perp$  thus {\em decreases exponentially} as a function of $N$.  This seems strange unless we remember that this can only be obtained at the expense of an {\em exponentially increasing} energy resource to carry out the computation, as evident from~(\ref{24}).

Other examples of this delicate balance between the energy required and the lower time limit for a well known quantum algorithm are given in Appendix B.
It is not that surprising that we could reduce the computation time with more energy resource, as the TSP is well parallelisable.  For example, in a parallel computation of $(N-1)!$ processing units, one for each of the permutations of $(2, \ldots, N)$ cities, we could solve the problem in efficient time but at the cost of a huge number of computer processors.

More interestingly and crucially, we could exploit the degree of freedom associated with $|\theta|$ and choose it as a function of $N$ such that
\begin{eqnarray}
(N-1)!|\theta|^{2N} &=& 1. \label{theta}
\end{eqnarray}
All the dependence on the size of the input has now been absorbed into $\theta = \theta(N)$ as an appropriately chosen function of $N$ as in~(\ref{theta}).  The parameter $\theta$ is an advantage of our proposed AQC algorithm that is denied or not evident elsewhere.  This choice would reduce {\em both} the lower time limit {\em and} energy resource to a manageable level of order ${\cal O}(1)$, independent of $N$, except the polynomial resources in $N$ needed for setting up the link, hooker and marker operators. 





We expect that the lower time limit~(\ref{time1}) could be saturated, as there is no reason to the contrary, and thus postulate that with quantum computation the TSP could be solved efficiently, both in energy resource and time complexity.  
Our time-energy consideration~(\ref{time1}), however, 
cannot give us the probability in obtaining the target ground state. 
For that we need to appeal to the quantum adiabatic theorem or otherwise.  

We hope to come back to a rigorous consideration of the issue of computational complexity elsewhere.  Nevertheless, we present in Appendix B below some further AQC illustrations in which the computational time complexity could be traded off for the energy resources.

\section{Concluding Remarks}
We have explicitly given an algorithm for the TSP in the context of AQC.  The algorithm may be implemented in the framework of quantum optics or quantum field theory or otherwise.  It may also inspire and lead to similar and appropriate algorithms for implementation with spin glass models.  It would be interesting if our algorithm could be `quantum digitised' for implementation on quantum computers with unitary gates acting on qubits.

It should be noted that similar algorithms could also be constructed with fermionic degrees of freedom, rather than with the bosonic operators $\{l\}$, $\{h\}$ and $\{m\}$ as in the above.  Fermionic number operators are more of a binary nature (with occupation number $n = 0,1$) but are perhaps more difficult to be implemented physically.

The algorithm is nonlocal.  And while nonlocality may be difficult to be physically implemented, it is the nonlocal and global characters, which are required for the TSP, that could give quantum computation, which have access to superposition and entanglement, the edge over its classical counterpart.

A full and rigorous consideration of the computational complexity is still needed for our algorithm, and we hope to address this issue elsewhere.  However, we have derived some estimates for a lower computing time limit and expect that this efficient limit could be saturated and met, in order of magnitudes, by the computation time derived from the quantum adiabatic theorem.

We also illustrate explicitly and emphasise here the important role of the energy resources required for physical computation (quantum or otherwise) in addition to the normally considered storage space and time resources.  This has been noticed previously~\cite{0204044, Kieu} in the context of some particular AQC algorithms.  But in general energy resource is an integral and essential component of computational complexity.  This is not peculiar to AQC but is a general feature, because quantum computation in general is a physical process after all, and any physical process does require energy in order to be unfolded in time --with the unfolding speed is usually an increasing function of energy and/or energy spreads~\cite{Kieu2}.

The author wishes to thank the referees for their careful consideration and insightful suggestions for the paper.

\appendix
\section{New class of time-energy uncertainty relations for time-dependent Hamiltonians}
In the paper~\cite{Kieu2} we obtain new class of time-energy uncertainty relations directly from the Schr\"odinger equation for time-dependent Hamiltonians in the general case.  Our derivation as well as the results are new and different to those in the existing literature.  It is important to note that only the initial state, and neither  the instantaneous eigenstates nor the full time-dependent wave function at any other times which would demand a full solution for a time-dependent Hamiltonian, is required for the time-energy relations.  This hallmark of our results in~\cite{Kieu2} enables their wider applicability and usefulness.





In particular, we also obtain some results for the adiabatic quantum computation AQC 
with time-varying Hamiltonian
${\cal H}_G (t)$ in the time interval $t\in[0,T]$ according to~(\ref{extrapolation})~and~(\ref{A5}),

We can set, without loss of generality, the initial ground state energy to zero to obtain the various {\em necessary} conditions
for the computing time $T^{AQC}$ at the end of the computation:
\begin{eqnarray}
2\hbar &\le& T^{AQC}_{\forall} \times \left(\int_0^1 g(\tau)d\tau\right)  \times\sqrt{\Delta_P E^2 +(E_P)^2}, \label{3} \\
\hbar\sqrt{2} &\le& T^{AQC}_{\perp} \times \left(\int_0^1 g(\tau)d\tau \right) \times{\Delta_P E}, \label{4}
\end{eqnarray}
here $E_P$ and and $\Delta_P E$ respectively are the expectation energy and the energy spread of the initial state $| g_I \rangle$ in terms of the target Hamiltonian $H_P$:
\begin{eqnarray}
E_P &\equiv& \langle g_I | H_P | g_I \rangle,\nonumber\\
\Delta_P E &\equiv& {\sqrt{\langle g_I|H_P^2|g_I\rangle -\langle g_I|H_P|g_I\rangle^2}}.
\label{spread}
\end{eqnarray}

The necessary conditions above can also be expressed differently but equivalently as that the system cannot fully explore the whole Hilbert space (that is, cannot reach certain dynamically allowable state) or evolve into an orthogonal state from the initial state if the evolution time is {\em less than}, respectively, the following AQC characteristic times:
\begin{eqnarray}
{\cal T}^{AQC}_{\forall} &\equiv& \frac{2\hbar}{\left(\int_0^1 g(\tau)d\tau\right) \times\sqrt{{\Delta_P E}^2 +(E_P)^2}}, \label{AQClowerbound}\nonumber\\
{\cal T}^{AQC}_{\perp} &\equiv& \frac{\hbar \sqrt 2}{\left(\int_0^1 g(\tau)d\tau\right) \times{\Delta_P E}}\label{AQCcharacteristic}
\end{eqnarray}
That is, if the computation time is less than ${\cal T}^{AQC}_{\forall}$ then there exists some state which is allowed by the dynamics but cannot be reached from the initial state.  And for evolution time less than ${\cal T}_{AQC\perp}$, the system cannot evolve to {\em any} state that is orthogonal to the initial state.  

The characteristic time in~(\ref{AQClowerbound}) could be considered as an estimate of the lower bound on the computing time; and as such, the more the energy and the more the spread of the initial state in energy with respect to the final Hamiltonian, the less the lower bound on computing time.  Note also that the inverses of these characteristic times are related to the measures of the interpolation rates of the AQC Hamiltonian~(\ref{extrapolation}); the slower the rates the higher the probabilities of ending the computation in the ground state of $H_P$.

Our characteristic computing time estimates for AQC depend only on the initial state of the computation, its energy expectation and also its energy spread as measured in terms of the final (observable) Hamiltonian of the computation. These estimates are not explicitly but only implicitly dependent on the instantaneous energy gaps at intermediate times of the spectral flow of the AQC time-dependent Hamiltonian.  

\section{Energy Resource as a Component of Computational Complexity}
We illustrate below the need for energy resources, not only in quantum computation but also in any physical computation, as an essential component for computational complexity, besides the usual resources of memory space and computing time.  We illustrate this point with the aid of the following unstructured search algorithms in AQC.

We first consider a quantum adiabatic algorithm~\cite{Roland2002, Wei} to locate the state $|m\rangle$ in a unsorted database set of normalised orthogonal states $\{|i\rangle, i = 1, \ldots, M\}$.  It is known that this algorithm has a computational complexity of ${\cal O}(\sqrt M)$ as that of Grover's algorithm~\cite{Grover}, a quadratic improvement on classical search.

For a AQC algorithm, we start with an initial state $|\phi_0\rangle$ that is a uniform superposition of all the states in the given search set,
\begin{eqnarray}
|\phi_0\rangle &\equiv& \sum_{i=1}^M c_i|i\rangle. \label{initial}
\end{eqnarray}
This state is the ground state of the initial Hamiltonian $H_0$,
\begin{eqnarray}
H_0 &=& {\bf 1} - |\phi_0\rangle \langle \phi_0|.
\end{eqnarray}
The target Hamiltonian $H_f$ is then designed to have the solution state $|m\rangle$ as the ground state,
\begin{eqnarray}
H_f &=& {\bf 1} - |m\rangle \langle m|.
\end{eqnarray}
The AQC is performed in the usual manner with a time-dependent Hamiltonian ${\cal H}_G (t)$ in the time interval $t\in[0,T]$ according to~(\ref{extrapolation})~and~(\ref{A5}),
\begin{eqnarray}
{\cal H}_G (t) &=& f(t/T)H_i + g(t/T)H_f. \label{H_G}
\end{eqnarray}

The energy expectation and energy spread of the target Hamiltonian as measured in the initial state $|\phi_0\rangle$ are, respectively,
\begin{eqnarray}
\langle \phi_0 | H_f |\phi_0 \rangle &=& 1 - |c_m|^2, \label{energy}
\end{eqnarray}
and
\begin{eqnarray}
\Delta_I E_f &\equiv& \sqrt{\langle \phi_0 | H_f^2 |\phi_0 \rangle - \langle \phi_0 | H_f |\phi_0 \rangle^2}, \nonumber\\
&=& \sqrt{|c_m|^2 - |c_m|^4}. \label{delta}
\end{eqnarray}

According to the time-energy uncertainty~(\ref{AQCcharacteristic}), the time estimate $T_\perp^{\rm search}$,
\begin{eqnarray}
T_\perp^{\rm search}&\sim& {\cal O} \left(\frac{1}{\int_0^1 g(\tau) d\tau \times \sqrt{|c_m|^2 - |c_m|^4}}\right), \label{time}
\end{eqnarray}
is the lower limit below which the initial state $|\phi_0\rangle$ {\em cannot} evolve into an orthogonal state under the dynamics governed by ${\cal H}_G(t)$ in~(\ref{extrapolation}).  This time limit is a typical measure of the computation time and thus should be of the same order of magnitude as the {\em best} AQC computation time,  as estimated according to the quantum adiabatic theorem, to obtain the target state $|m\rangle$ with reasonable probability.

For the case of the initial state is a {\em uniform} superposition of all the states, that is,
\begin{eqnarray}
c_i &=& \frac{1}{\sqrt M}, \forall i = 1, \ldots, M,
\end{eqnarray}
we then have from~(\ref{time})
\begin{eqnarray}
T_\perp^{\rm search} &\stackrel{M\gg 1}{\sim}& {\cal O} \left(\frac{\sqrt M}{\int_0^1 g(\tau) d\tau}\right). \label{A8}
\end{eqnarray}
This time estimate, with $g$ for which $ \int_0^1 g(\tau) d\tau\sim {\cal O}(1)$, is indeed of the same order of magnitude as the time complexity ${\cal O}(\sqrt[]{M})$ for the AQC~\cite{Roland2002} as normally obtained from the energy gap of the two lowest eigenvalues in the spectral flow of ${\cal H}_G(t)$ according to the quantum adiabatic theorem.

In contrast to those derived from the quantum adiabatic theorem, the time estimate $T_\perp^{\rm search}$ here depends only on the extrapolating function $g$, the initial state and the target Hamiltonian.  Our lower bound estimate, furthermore for this particular algorithm,  is independent of all other amplitudes $c_i$ for $i\not= m$. It depends only on the coefficient $c_m$ of the target state in the superposition~(\ref{initial}).  We thus could improve on the time ${\cal O}(\sqrt{M})$ if we have some information that leads to higher priori probability for the target state $|m\rangle$, such that $|c_m| > 1/\sqrt{M}$.  

In addition to that, we could also exploit the extra degree of freedom of the extrapolating function $g$ to reduce the time estimate~(\ref{A8}).  For example, with the choice
\begin{eqnarray}
g(\tau) &\longrightarrow& \sqrt{M}g(\tau) \label{A10}
\end{eqnarray}
substituting in~(\ref{A8}) we could have reduced the lower time limit $T_\perp^{\rm search} \sim {\cal O}(1)$!  This choice and its computation time have also been confirmed in~\cite{Wei}.

As another example, the authors of~\cite{0204044} employ a different function $g(\tau)$ but which also grows with $\sqrt M$,
\begin{eqnarray}
g(\tau) &\longrightarrow& \tau + \sqrt{M}\tau(1-\tau) \label{A11}.
\end{eqnarray}
This once again reduces the computation time to ${\cal O}(1)$, also in agreement with~(\ref{A8}) whence
\begin{eqnarray}
\int^1_0 g(\tau) d\tau &\longrightarrow& 1/6 + \sqrt{M}/2.
\end{eqnarray}

All of the above reductions for $T_\perp^{\rm search}$ match, in orders of magnitude, the time complexities derived in~\cite{0204044, Wei} directly from a consideration of spectral-flow energy gap according to the quantum adiabatic theorem.  This agreement is remarkable, as our results above are not derived directly from the quantum adiabatic theorem but from a general consideration of time-energy uncertainty relation for time-dependent Hamiltonians~\cite{Kieu2}.

Such an agreement, however, is not unexpected.  It should be reminded again here that our time measure $T_\perp^{\rm search}$ is a necessary lower limit in the sense that if the computation time is less than that then the initial state cannot evolve into an orthogonal state.  But longer computation time, $T > T_\perp^{\rm search}$, is {\em not} a sufficient condition; for sufficiency we would need to involve the quantum adiabatic theorem.  However, the simply calculated time measure $T_\perp^{\rm search}$ does agree in order of magnitudes with the estimate of the computational time more comprehensively derived.  This agreement of our results and those in~\cite{Wei} demonstrates that these necessary lower limits can in fact be saturated in this case by judicious choice of the extrapolating functions $f$ and $g$~\cite{Roland2002, Wei}.

More importantly, we want to point out and emphasise here that although we may be able to reduce the time complexity to ${\cal O}(1)$, as with the choice of~(\ref{A10})or~(\ref{A11}), we need to consider also the energy resource required for the computation.  The choice of~(\ref{A10}) can, in fact, only be had at the cost of an increase in the energy required:
\begin{eqnarray}
\max_{0\le t \le T} \langle \phi_0| {\cal H}_G(t) |\phi_0\rangle &\to& {\cal O}(\sqrt M).
\end{eqnarray}
That is, a reduction in time complexity (from ${\cal O}(\sqrt M)$ to ${\cal O}(1)$) incurs and is balanced by an increase in the cost in energy resource (from ${\cal O}(1)$ to ${\cal O}(\sqrt M)$).  That is, in general the computational complexity of the AQC~(\ref{H_G}) is of the order ${\cal O}(\sqrt M)$, taking into account both the energy and time resources.

One could go to the extreme and implement the choice 
\begin{eqnarray}
g(\tau) &\longrightarrow& {\rm e}^Mg(\tau) \label{A12}
\end{eqnarray}
in order to have an exponentially {\em decreasing} computation time with $M$.  But of course we then at the same time have to pay exponentially for the energy required for the physical AQC computation,
\begin{eqnarray}
\max_{0\le t \le T} \langle \phi_0| {\cal H}_G(t) |\phi_0\rangle &\to& {\cal O}({\rm e}^M).
\end{eqnarray}

{\em The message here is that in considering the computational complexity in general we need to consider also the energy resources in addition to the usually considered space and time complexity.}  

See~\cite{0110020, 0204087, 0205048, 0208135, 0309201} for other AQC Hamiltonians for this search problem.  For these and more general time-dependent Hamiltonians $H(\tau)$, which do not assume the particular form of~(\ref{extrapolation})~and~(\ref{A5}), we still could estimate some lower bounds on the computational time through the general uncertainty relations derived in~\cite{Kieu2}:
\begin{eqnarray}
2\hbar &\le& \int^{T_\forall}_0{\left\|H(\tau)|\phi_0\rangle\right\|}d\tau,  \label{parallell}
\end{eqnarray}
and
\begin{eqnarray}
\hbar\sqrt{2} &\le& \int^{T_\perp}_0{\left\|(H(\tau) -\beta(\tau))|\phi_0\rangle\right\|}d\tau , \label{perpendicular}
\end{eqnarray}
where $\beta(\tau)$ is an arbitrary function, whose form could be chosen to facilitate tight inequalities.

\bibliography{TSP_ib}

\end{document}